# Field-free perpendicular magnetization switching of low critical current density at room temperature in TaIrTe$_4$/ferromagnet heterostructures

Lujun Wei[1, 2, a)], Pai Liu[1], Jincheng Peng[1], Yanghui Li[1], Lina Chen[1], Ping Liu[1], Feng Li[1], Wei Niu[1], Fei Huang[1], Jiaju Yang[1], Shuang Zhou[1], Yu Lu[3], Tianyu Liu[3], Jiarui Chen[3], Weihao Wang[3], Jian Zhang[3], Jun Du[3, 4, a)], Yong Pu[1, a)]

[1]*School of Science, Nanjing University of Posts and Telecommunications, Nanjing 210023, China*

[2]*National Laboratory of Solid State Microstructures, Nanjing University, Nanjing 210093, China*

[3]*Department of Physics, Nanjing University, Nanjing 210093, China*

[4]*National Key Laboratory of Spintronics, Nanjing University, Suzhou 215163, China*

[a)]Authors to whom correspondence should be addressed: wlj@njupt.edu.cn; jdu@nju.edu.cn; and yongpu@njupt.edu.cn.

## ABSTRACT

Spin-orbit torque-induced perpendicular magnetization switching has attracted much attention due to the advantages of nonvolatility, high density, infinite read/write counts, and low power consumption in spintronic applications. To achieve field-free deterministic switching of perpendicular magnetization, additional magnetic field, magnetic layer assistance, or artificially designed structural symmetry breaking are usually required, which are not conducive to the high-density integration and application of low-power devices. However, 2D type-II Weyl semimetals with low-symmetry structures have recently been found to generate $z$-spin-polarized currents, which may induce out-of-plane damping-like torques to their neighboring ferromagnetic layers, and realize deterministic perpendicular magnetization switching at zero magnetic field. In this Letter, we report that current-induced field-free magnetization switching at room temperature can be achieved in a perpendicularly magnetized TaIrTe$_4$/Pt/Co/Pt device, and the critical switching current density can be lowered to be about $2.64\times10^5$ A·cm$^{-2}$. When the current is applied along the low-symmetry TaIrTe$_4$ $a$-axis, the out-of-plane and in-plane spin Hall conductivities are estimated to be $1.00\times10^4\times\hbar/2e$ $(\Omega\cdot m)^{-1}$ and $1.20\times10^4\times\hbar/2e$ $(\Omega\cdot m)^{-1}$, respectively. This study suggests that TaIrTe$_4$ has great potential for the design of room-temperature efficient spintronic devices.

Spin-orbit torque (SOT) has shown great potential in driving the magnetization switching of ferromagnets, providing new opportunities for the development of energy-efficient and high-density spintronic devices.[1,2] Conventional heavy metal (HM)/ferromagnetic (FM) heterostructures usually require the assistance of external magnetic field when utilizing SOTs to achieve a deterministic switching of perpendicular magnetization.[3,4] To address this limitation, researchers have proposed various means, such as constructing asymmetric structures, interlayer/exchange coupling, and utilizing additional electric field modulation, to achieve field-free perpendicular magnetization switching.[5-10] Although these methods are effective, they tend to involve complex fabrication processes or require additional magnetic layers, which limits the high-density integration and application of low-power devices.[1,11-13]

The out-of-plane damping-like SOT provides a more efficient solution, which counteracts the magnetic damping, thus enabling the switching of perpendicularly magnetized ferromagnetic materials at zero magnetic field and with lower power consumption.[14-17] Recent studies have shown that the type-II Wely semimetals with non-centrosymmetric structures (e.g., $WTe_2$,[18-31] $TaIrTe_4$,[11,32-35] etc.) exhibit a strong spin-orbit coupling, capable of generating z-spin-polarized ($\sigma_z$) currents and exerting the out-of-plane SOT ($\tau_z \sim \mathbf{m}\times(\mathbf{m}\times\boldsymbol{\sigma_z})$) to the neighboring ferromagnet, thus achieving a field-free perpendicular magnetization switching. However, these 2D materials are susceptible to oxidation and further degradation at room temperature.[36-38] Therefore, researchers usually employ magnetron sputtering to deposit FM layers on thin flakes of 2D materials exfoliated in a vacuum or in an inert gas environment to construct nonmagnetic (NM)/FM heterostructures.[11,32] The NM/FM interface has a significant effect on spin transport.[39-42] Therefore, this approach may lead to a degradation of the quality of the interface between 2D materials and FM layers.

In this Letter, we are devoted to the preparation of high-quality conventional ferromagnets interfaced with the Wely-II semimetal TaIrTe$_4$ to explore properties of its SOTs and current-induced magnetization switching. Specifically, we first prepared Pt/Co/Pt trilayers by magnetron sputtering on SiO$_2$/Si substrates, and then successfully constructed clean TaIrTe$_4$/FM interfaces by exfoliating TaIrTe$_4$ and transferring it onto the Pt/Co/Pt trilayers in a nitrogen-filled glove box. We investigated the out-of-plane and in-plane SOTs of the TaIrTe$_4$/Pt/Co/Pt device by measuring anomalous Hall resistance (AHR) loop shift and first/second harmonic Hall voltages. We observed the field-free current-driven magnetization switching at room temperature with a low critical switching current density of about $2.64 \times 10^5$ A·cm$^{-2}$. This study provides an important reference value for the low-power-consumption spintronic devices by utilizing the two-dimensional topological materials.

Firstly, an Pt(2 nm)/Co(0.72 nm)/Pt(2 nm) trilayer was deposited on a clean SiO$_2$/Si substrate by magnetron sputtering. Next, the trilayer's Hall bar sample with widths of 5 or 8 μm was prepared by electron beam lithography and ion etching. Subsequently, the mechanically exfoliated long strip of TaIrTe$_4$ flake was transferred onto the FM-layered Hall bar by a transfer platform, ensuring that the long edge of the TaIrTe$_4$ flake was parallel to the long edge of the Hall bar, as shown in Fig. (a). These TaIrTe$_4$ flakes were obtained by mechanically exfoliating a single crystal of T$_d$-TaIrTe$_4$ (HQ Graphene Ltd.). Finally, the TaIrTe$_4$ flake was covered with an h-BN flake to prevent it from being oxidized.

In order to study the role of the TaIrTe$_4$ flakes, we first measured the first/second harmonic Hall voltages and current-driven magnetization switching curves for the Pt/Co/Pt Hall bar sample. Afterwards, the same tests were carried out after transferring the TaIrTe$_4$ flakes onto the Hall bar. The thickness of the TaIrTe$_4$ flake was determined to be about 36 nm by an atomic force microscopy (AFM, Dimension ICON), as shown in Fig. S1 of the supplementary material. Current-induced magnetization switching curves, AHR curves, and first/second-harmonic Hall measurements were performed in a home-made magneto-electric transport system equipped with a soucemeter (Keithley 6221), a nanovoltage (Keithley 2182) and a lock-in amplifier (Stanford 830). For the harmonic Hall measurement, a Keithley 6221 was used to apply a sinusoidal a.c. current with a fixed frequency of 131.0 Hz, and the output voltage was measured by a lock-in amplifier. The out-of-plane magnetic hysteresis (*M–H*) loop of the Pt/Co/Pt films was measured by a vibrating sample magnetometer (VSM, Microsense EV7). All the measurements were performed at room temperature.

The stacking structure of the device is schematically shown in Fig. 1(b). The structure of the $T_d$-phase TaIrTe$_4$ exhibits mirror symmetry with respect to the *bc*-plane, but not with respect to the *ac*-plane,[43,44] suggesting significant structural anisotropy in the device. In experiments, polarized Raman spectra are usually used to distinguish the axis of TaIrTe$_4$ flakes.[11,34,45,46] By using a Raman spectrometer (Renishaw inVia) with a wavelength of 633 nm, the Raman spectra of the TaIrTe$_4$ flake were obtained at various polarization angles, as shown in Fig. 1(c), and the peak positions are consistent with those reported previously.[45] The relationship between the intensity of the peak at ~147 cm$^{-1}$ and the polarization angle is shown in Fig. 1(d). The intensity maxima locates at the *a*-axis,[45] indicating that the 0°- and 90°-direction correspond to the *a*-axis and *b*-axis of the TaIrTe$_4$ flake, respectively, agreeing well with the long and short edge of the TaIrTe$_4$ flake in the device [Fig. 1(a)].

Figure 2(a) demonstrates the out-of-plane AHR curve for the Pt/Co/Pt trilayer (FM sample), and the saturation value of AHR resistance was obtained to be 1.05 Ω. Subsequently, we measured its in-plane AHR curve, and the result is displayed in Fig.

2(b). By fitting the low-field in-plane AHR curve [see Fig. S2 of the supplementary material], the perpendicularly magnetic anisotropy (PMA) field is obtained to be about 3500 Oe. Figures 2(c) and (d) show the current-driven magnetization switching curves for the FM sample and the TaIrTe$_4$/FM device with $I//a$-axis, respectively. It can be clearly seen that no deterministic switching was observed at zero magnetic field for the FM sample [Fig. 2(b)] and the TaIrTe$_4$/FM device with $I//b$-axis [Fig. S3 of the supplementary material]. However, when a current was applied along the $a$-axis, deterministic magnetization switching could be achieved at zero field [Fig. 2(d)] with a switching ratio of about 28 %. Here, the switching ratio is defined as the ratio of the AHR to the saturated AHR.

The critical switching current density of the device with $I//a$-axis at 0 Oe is about $2.64\times10^5$ A·cm$^{-2}$ [see Note S1 of the supplementary material for details], which is two orders of magnitude lower than that of the conventional HM/FM structures.[47,48] In addition, this critical current density is about one order of magnitude less than that of TaIrTe$_4$/Ti/CoFeB/MgO device[11,32] (~ $2.4\times10^6$ A·cm$^{-2}$) and van der Waals heterojunction devices (e.g., WTe$_2$/Fe$_3$GeTe$_2$[24,25] (~ $3.5\times10^6$ A·cm$^{-2}$), WTe$_2$/Fe$_3$GaTe$_2$[29] (~ $2.23\times10^6$ A·cm$^{-2}$), and TaIrTe$_4$/Fe$_3$GaTe$_2$[33] (~ $2.56\times10^6$ A·cm$^{-2}$)). This value is also smaller compared to the critical current densities of topological insulators (e.g., (Bi$_{1-x}$Sb$_x$)$_2$Te$_3$ (~ $(0.52-2.43)\times10^6$ A·cm$^{-2}$),[48] Bi$_2$Te$_3$ (~ $2.2\times10^6$ A·cm$^{-2}$),[49] etc.) as spin-source materials switching neighboring perpendicular ferromagnets.

For the Pt/Co/Pt sample, the deterministic magnetization switching occurs under the assistance of external magnetic fields, and the switching polarity is opposite under positive and negative magnetic fields, as shown in Fig. 2(c). However, the switching curve of the TaIrTe$_4$/Pt/Co/Pt device with $I//a$-axis disappears at a magnetic field of about -10 Oe [Fig. 2(d)], which may be induced by the effect of out-of-plane SOTs, and this is similar to current-induced magnetization switching curves that have been reported for the Mn$_3$Sn/[Co/Ni]$_2$[17] and the WTe$_2$/Pt/Co/Pt[28] systems.

To further confirm the existence of *z*-spin-polarized current for the TaIrTe$_4$/Pt/Co/Pt device, we performed AHR loops measurement under different d.c. currents, as shown in Fig. 3. When a current of ±1 mA was applied along the *a*-axis, AHR loop shift could not be observed [Fig. 3(a)]. By applying a larger current of ±8 mA along the *a*-axis, a significant AHR loop shift is observed [Fig. 3(b)]. When a current of 8 mA or -8 mA was applied, the AHR loop of $R_H$ vs $H$ shifted towards the positive or negative magnetic field direction, respectively, confirming the existence of *z*-spin polarization.

The out-of-plane effective field ($H_z$) values generated by the AHR loops measured at different d.c. currents are summarized in Fig. 3(d). $H_z$ is defined as $H_z = [H_{shift}(+I) - H_{shift}(-I)]/2$, where $H_{shift}(-I)$ and $H_{shift}(+I)$ are the offsets of AHR loops at negative and positive currents, respectively. At 8 mA, the $H_z$ is about 5 Oe, which is used to estimate the out-of-plane spin conductivity ($\sigma_S^z$) by the equation[27]

$$\sigma_S^z = \frac{\mu_0 M_S t_{FM} H_z}{E}. \tag{1}$$

Here, $\mu_0$, $M_S$, $t_{FM}$ and $E$ are the vacuum permeability, the saturation magnetization of the FM layer, the thickness of the Pt/Co/Pt trilayer ($t_{FM}$ = 4.72 nm) and the electric field applied to the device. The $M_s$ value of the 4.72 nm-thick Pt/Co/Pt trilayer is 132.1 emu/cm$^3$ [see Fig. S4 of the supplementary material]. $E$ is about 90.16 kV/cm obtained by using the equation $E = IR_{xx}/d$, where $I$, $R_{xx}$ and $d$ are applied current, the longitudinal resistance and thickness of the device, respectively. By substituting these parameters of Eq. (1), the out-of-plane spin conductivity of TaIrTe$_4$ *a*-axis is obtained as $\sigma_S^z = 1.00 \times 10^4 \times \hbar/2e \ (\Omega \cdot m)^{-1}$. Here, the $\hbar$ and $e$ are the reduced Planck constant and the electron charge, respectively. The charge resistivity $\rho_{TaIrTe_4}$ of TaIrTe$_4$ flakes is about 4.44×10$^{-6}$ Ω·m measured by the four-probe transport measurements with the current flowing along the *a*-axis. By using $\xi_{sh}^z = \rho_{TaIrTe_4}\sigma_s^z$, the out-of-plane charge-to-spin conversion efficiency is estimated to be $\xi_{sh}^z = 0.044$, which is close to that in previous reports.[11,32] However, when the current is along the *b*-axis of the device, there is no shift of the AHR loops shown in Figs. 3(c) and (d), suggesting almost zero out-of-plane SOTs' production.

In addition, the first and second harmonic Hall measurements were performed to determine the effective fields of $H_{DL}$ and $H_{FL}$ induced by in-plane damping-like torque and field-like torque, respectively, in both the TaIrTe$_4$/FM device and the FM sample. In the experiments, we applied an a.c. current to the devices along the *x*-direction (*y*-direction) to measure the first harmonic voltage ($V_{1\omega}$) and the second harmonic voltage ($V_{2\omega}$) along the *y*-direction (*x*-direction), respectively.

The external in-plane magnetic field applied during the measurement is much smaller than PMA field of the FM sample, and the direction of the FM layer' magnetization is along the *z* or -*z* axis. In this case, the polarized direction of the *z*-spin-polarized current is colinear with the direction of the magnetization, so the out-of-plane spin $\sigma_z$-induced damping-like SOT is almost zero. In this case, measuring signals are mainly contributed from the effective field generated by the $\sigma_y$-spin-induced SOTs on the magnetization. The $H_{DL}$ and $H_{FL}$ are estimated by using the equation,[50,51]

$$H_{DL(FL)} = \frac{-2\frac{\partial V_{2\omega}}{\partial H_{x(y)}}}{\frac{\partial^2 V_\omega}{\partial H_{x(y)}^2}}, \tag{2}$$

where $H_x$ and $H_y$ correspond to the applied magnetic field along the *x* and *y* direction, respectively. Figures 4(a) and (b) show the measured and fitted curves of $V_{1\omega}$ and $V_{2\omega}$ vs magnetic field (*H*) at ±$M_z$ for the device with *I* = 3.5 mA along the *a*-axis, respectively. The fitted results for the device with different a.c. currents along the *a*-axis and *b*-axis, and the FM sample are shown in Figs. S5-S6, S7-S8 and S9-S10 in the supplementary material.

Figures 4(c) and (d) show the dependence of the averaged effective field values of $\bar{H}_{DL}$ and $\bar{H}_{FL}$ on the applied current. The $\bar{H}_{DL}$ and $\bar{H}_{FL}$ are define as $\bar{H}_{DL}$ = [$H_{DL}(+M_z)$-$H_{DL}(-M_z)$]/2 and $\bar{H}_{FL}$ = [$H_{FL}(+M_z)$-$H_{FL}(-M_z)$]/2. Here, the values of $H_{DL}(+M_z)$/$H_{FL}(+M_z)$ and $H_{DL}(-M_z)$/$H_{FL}(-M_z)$ are obtained by fitting the first/second harmonic Hall curves at +$M_z$ and -$M_z$, respectively. Subsequently, we performed a linear fit according to the $\bar{H}_{DL} \sim I$ data and obtained the in-plane damping-like torque effective field per unit current [Fig. 4(c)] for the Pt/Co/Pt sample, the TaIrTe$_4$/Pt/Co/Pt device with current applied along the *a*-axis and the *b*-axis to be as 0.16 Oe·mA$^{-1}$, 0.42 Oe·mA$^{-1}$ and 1.54 Oe·mA$^{-1}$, respectively. The field-like torque effective fields per unit current [Fig. 4(d)] for those device are estimated to be 0.08 Oe·mA$^{-1}$, 0.61 Oe·mA$^{-1}$ and 0.50 Oe·mA$^{-1}$, respectively.

The in-plane spin Hall conductivity of the Pt/Co/Pt sample is calculated to be $\sigma_s^y(FM)$ = 0.12×10$^4$×$\hbar/2e$ $(\Omega \cdot m)^{-1}$ according to equation (1). When the current is applied along the *a*-axis of the TaIrTe$_4$, the in-plane spin Hall conductivity of the TaIrTe$_4$/Pt/Co/Pt device can be estimated as $\sigma_s^y$(*a*-axis) = 1.20×10$^4$×$\hbar/2e$ $(\Omega \cdot m)^{-1}$, which is comparable to (5.437 ± 0.043)×10$^4$×$\hbar/2e$ $(\Omega \cdot m)^{-1}$ reported previously on the TaIrTe$_4$/Py bilayers measured by spin-torque ferromagnetic resonance (ST-FMR) [11] and is one order of magnitude larger than that of the Pt/Co/Pt sample. This suggests that the main contribution to $\sigma_s^y$(*a*-axis) for the device comes from the TaIrTe$_4$ layer. We estimated the charge-to-spin conversion efficiency of the device as $\xi_{sh}^y$ = 0.053, which is consistent with those reported previously.[11,32] The in-plane spin Hall conductivity of the TaIrTe$_4$/Pt/Co/Pt device with *I*//*b*-axis is estimated to be 0.31×10$^4$×$\hbar/2e$ $(\Omega \cdot m)^{-1}$, which is much smaller than that of the device with *I*//*a*-axis.

In conclusion, the TaIrTe$_4$/Pt/Co/Pt device was constructed, and the current-induced field-free deterministic switching of perpendicular magnetization was realized at room temperature, and the critical switching density was 2.64×10$^5$ A·cm$^{-2}$. When the current is applied along the *a*-axis of the device, the out-of-plane and in-plane spin Hall conductivities of the TaIrTe$_4$ flake are about $1.00×10^4× \hbar/2e\ (\Omega \cdot m)^{-1}$ and $1.20×10^4×\hbar/2e\ (\Omega \cdot m)^{-1}$, respectively, and the out-of-plane and in-plane charge-to-spin conversion efficiencies are about 0.044 and 0.053, respectively. The present study shows the TaIrTe$_4$/FM device has a large out-of-plane spin Hall conductivity and a low critical current density, which provides a reference value for the application of low-power-consumption spintronics devices.

Supplementary material

See the supplementary material for detailed information of the following: the AFM image of the TaIrTe$_4$ flake in the device, current-induced magnetization switching curves of the TaIrTe$_4$/FM device with *I*//*b*-axis, fitted PMA field of the FM sample, out-of-plane *M-H* loop of the Pt/Co/Pt trilayer, experimental and fitted $V_{1\omega}$ and $V_{2\omega}$ for the device and the FM sample, and estimation of critical current density with the parallel current model.

This work was supported by the National Key R&D Program of China (Nos. 2023YFA1406603 and 2022YFA1403602), the National Natural Science Foundation of China (Nos. T2394473, 12374112, 52001169, and 61874060), and Open Foundation from Anhui key laboratory of low-energy quantum materials and devices.

AUTHOR DECLARATIONS

Conflict of Interest

The authors have no conflicts to disclose.

DATA AVAILABILITY

The data that support the findings of this study are available within the article and its supplementary material.

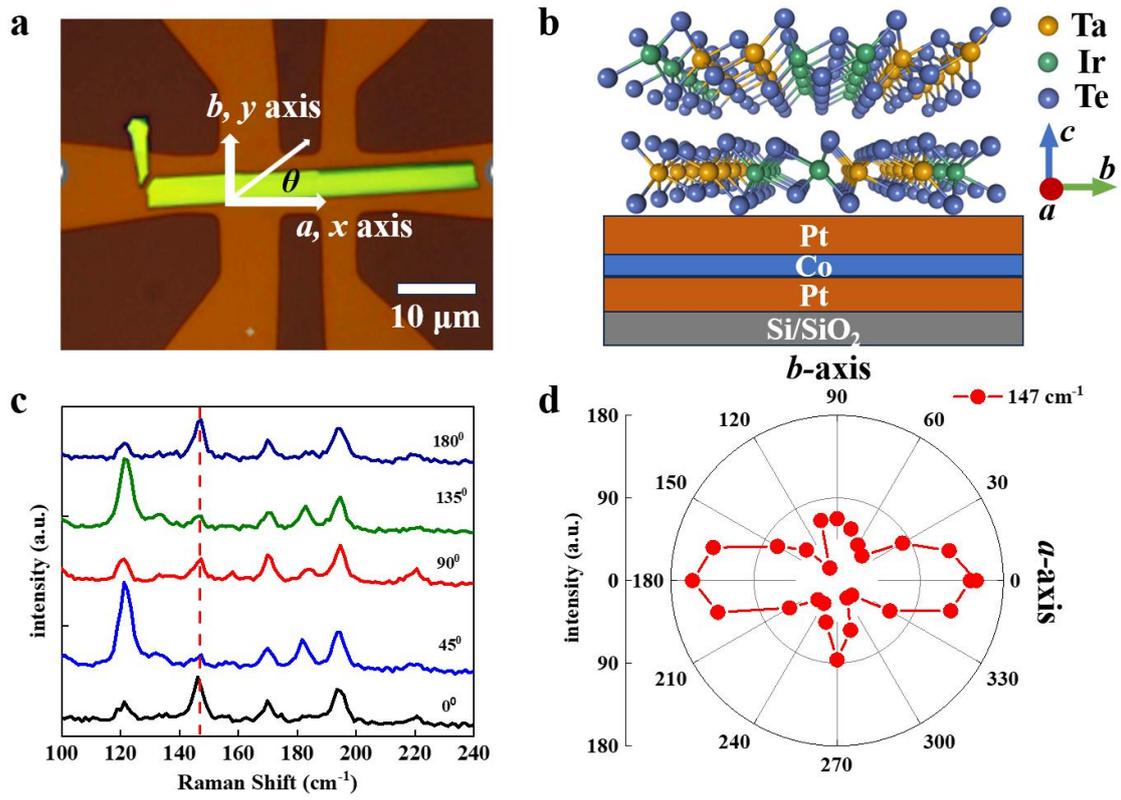

**FIG. 1.** Optical photograph (a) and schematic stacking (b) of the device. (c) Raman spectra of the TaIrTe$_4$ flake collected at different polarization angles ($\theta$) defined in (a). (d) Raman spectral intensity versus polarization angle ($\theta$) for the TaIrTe$_4$ flake at 147 cm$^{-1}$.

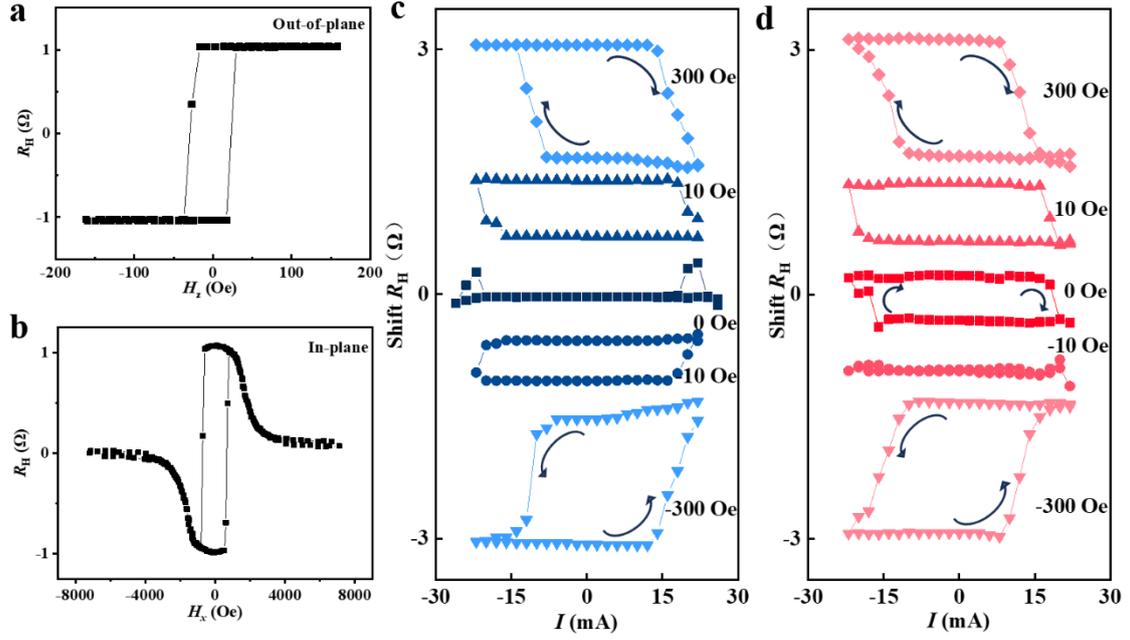

**FIG. 2.** Out-of-plane (a) and in-plane (b) AHR curves of the Pt/Co/Pt sample. Current-induced magnetization switching loops of the Pt/Co/Pt trilayer (c) and the TaIrTe$_4$/Pt/Co/Pt devices (d) with the current applied along the *a*-axis under different external magnetic fields applied in the current direction. The loops are manually shifted vertically for better visualization.

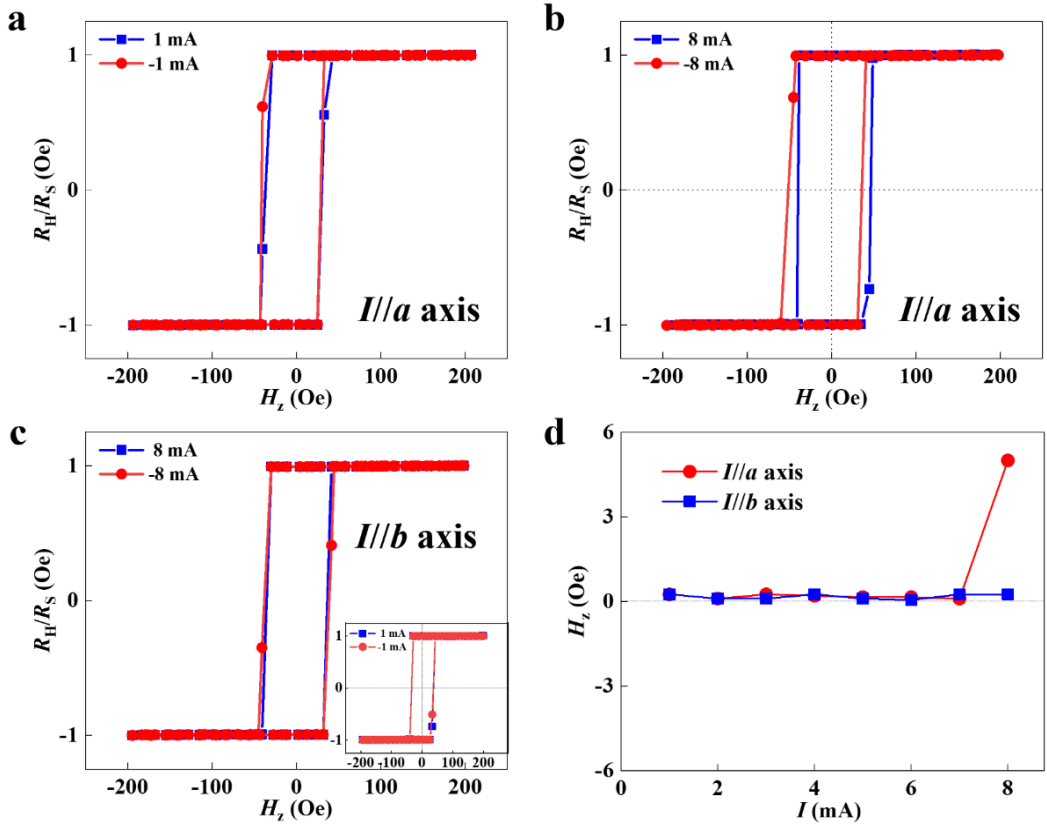

**FIG. 3.** AHR loops measured by applying ±1 mA (a) and ±8 mA (b) along the *a*-axis of the device. (c) AHR loops measured by applying ±8 mA and ±1 mA (the insert) along the *b*-axis of the device. (d) The out-of-plane effective field $H_z$ under different applied current.

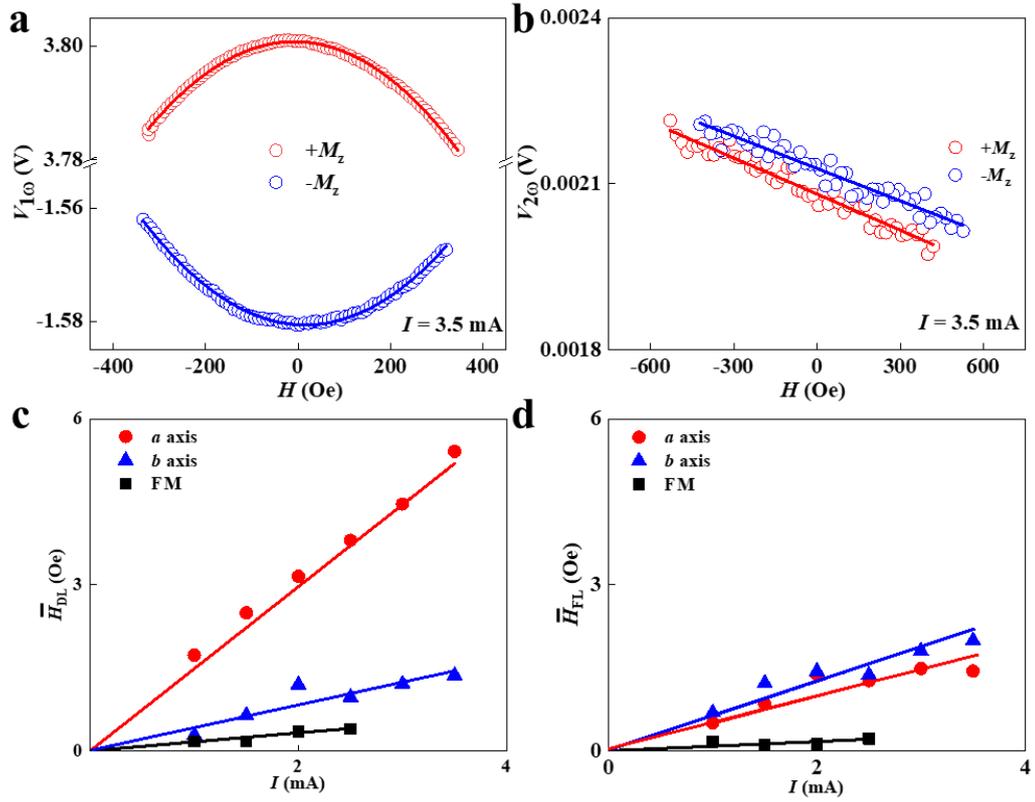

**FIG. 4.** $V_{1\omega}$ (a) and $V_{2\omega}$ (b) versus $H$ measured by applying a current of 3.5 mA along the *x*-direction (i.e. *a*-axis) of the TaIrTe$_4$/Pt/Co/Pt device. The solid lines show the fitted results. The current dependence on the $\bar{H}_{DL}$ (c) and $\bar{H}_{FL}$ (d) for the TaIrTe$_4$/Pt/Co/Pt device with $I$ applied along the *a*-axis and *b*-axis and for the FM sample, respectively.